\documentclass[sigconf]{acmart}

\usepackage{booktabs} % For formal tables
\usepackage{nicefrac}  

\newcommand{\Prob}{{P}}

\setcopyright{rightsretained}
\acmDOI{}

\acmISBN{}

\acmConference[VALUETOOLS'17]{11th EAI International Conference on Performance Evaluation Methodologies and Tools}{December 2017}{Venezia, Italy} 
\acmYear{2017}
\copyrightyear{2017}

\acmArticle{}
\acmPrice{}

\editor{}
\editor{}
\editor{}

\begin{document}
\title{Trading Optimality for Performance in Location Privacy}
%\titlenote{This work is partially supported by a Digiteo-DigiCosme grant.}
\subtitle{Extended Abstract}
%\subtitlenote{The full version of the author's guide is available as
%  \texttt{acmart.pdf} document}

\author{Konstantinos Chatzikokolakis}
%\authornote{}
%\orcid{}
\affiliation{%
  \institution{CNRS and Ecole Polytechnique}
%  \streetaddress{}
%  \city{} 
%  \state{} 
%  \postcode{}
}
%\email{}

\author{Serge Haddad}
%\authornote{}
%\orcid{}
\affiliation{%
  \institution{ENS Cachan}
%  \streetaddress{}
%  \city{} 
%  \state{} 
%  \postcode{}
}
%\email{}

\author{Ali Kassem}
%\authornote{}
%\orcid{}
\affiliation{%
  \institution{INRIA and Ecole Polytechnique}
%  \streetaddress{}
%  \city{} 
%  \state{} 
%  \postcode{}
}
%\email{}

\author{Catuscia Palamidessi}
%\authornote{}
%\orcid{}
\affiliation{%
  \institution{INRIA and Ecole Polytechnique}
%  \streetaddress{}
%  \city{} 
%  \state{} 
%  \postcode{}
}
%\email{}

%\author{Tymofii Prokopenko}
%%\authornote{}
%%\orcid{}
%\affiliation{%
%  \institution{INRIA and Ecole Polytechnique}
%%  \streetaddress{}
%%  \city{} 
%%  \state{} 
%%  \postcode{}
%}
%%\email{}

\renewcommand{\shortauthors}{K. Chatzicocolakis et al.}

%\begin{abstract}
%This paper provides a sample of a \LaTeX\ document which conforms,
%somewhat loosely, to the formatting guidelines for
%ACM SIG Proceedings.\footnote{This is an abstract footnote}
%\end{abstract}

%\begin{CCSXML}
%<ccs2012>
%  <concept_id>10010520.10010553.10010562</concept_id>
%  <concept_desc>Computer systems organization~Embedded systems</concept_desc>
%  <concept_significance>500</concept_significance>
%  <concept_id>10010520.10010575.10010755</concept_id>
%  <concept_desc>Computer systems organization~Redundancy</concept_desc>
%  <concept_significance>300</concept_significance>
%  <concept_id>10010520.10010553.10010554</concept_id>
%  <concept_desc>Computer systems organization~Robotics</concept_desc>
%  <concept_significance>100</concept_significance>
%  <concept_id>10003033.10003083.10003095</concept_id>
%  <concept_desc>Networks~Network reliability</concept_desc>
%  <concept_significance>100</concept_significance>
%</ccs2012>  
%\end{CCSXML}
%
%\ccsdesc[500]{Computer systems organization~Embedded systems}
%\keywords{ACM proceedings, \LaTeX, text tagging}

\maketitle

\section{Introduction}
Location-Based Services (LBSs) provide invaluable assistance in our everyday activities, however they also pose serious threats to our privacy. Location data can, in fact, expose sensitive aspects of the user's private life, see for instance  \cite{Freudiger:11:FC}. 
There is, therefore, a growing interest  in the development of mechanisms to protect location privacy  during the use of LBSs. Most of the approaches in the literature are based on perturbing the user's  location, see, for instance, \cite{Andres:13:CCS,Shokri:17:TPS,Chatzikokolakis:17:POPETS,Oya:17:CORR}.  Obviously, the perturbation must be done with care, in order to preserve the utility of the service. 

Nowadays, the most popular methods (including all those mentioned above)  are probabilistic, in the sense that the perturbation is done by adding noise according to some probability distribution. Indeed, it is generally recognized that probabilistic mechanisms offer a better trade-off between privacy and utility. In this abstract we focus on the approach proposed in \cite{Shokri:17:TPS}, which achieves an optimal trade-off by using linear optimization techniques. The idea is to express the desired level of privacy in the form of linear constraints, and the utility as the objective (linear) function to optimize.\footnote{In  \cite{Shokri:17:TPS} 
 the authors fix the utility and optimize  privacy. We do the reverse as our notion of privacy can only be expressed as a set of constraints, not as an objective function.} The variables of the linear program are the conditional probabilities of reporting a location $y$ when the real one is $x$, and their values, once computed, completely define the mechanism. 

We consider the notion of privacy proposed in \cite{Andres:13:CCS}, called \emph{geo-indistinguishability}. A mechanism provides 
geo-indistinguishability if the probability of reporting a location  $y$ when the real location is $x$ is ``almost the same'' as that of every other location $x'$ at a distance $d(x,x')$ from $x$, where ``almost the same'' means that the ratio of the probabilities is bound by $\exp(\varepsilon \cdot d(x,x'))$, with 
$\varepsilon$ being the level of privacy we want to obtain per unit of distance. Formally: 
\begin{eqnarray}\label{eqn:geo-ind}
\Prob(y\mid x) \leq \exp(\varepsilon \cdot d(x,x')) \cdot \Prob(y\mid x')
\end{eqnarray}
Intuitively, this means that $x$ is ``$\varepsilon  \cdot \ell$-indistinguishable''   from the other locations $x'$  which are at distance at most $\ell$ from $x$, where $\varepsilon$ represents the level of indistinguishability that we want to achieve per unit distance. As explained in \cite{Andres:13:CCS},  geo-indistinguishability is based on (an extended form of) \emph{differential privacy} \cite{Dwork:06:TCC}, and it inherits its appealing properties. Notably, the robustness with respect to composition, the independence from the prior, and a natural interpretation in terms of Bayes adversaries. 

For the utility loss $\mathcal{U}$ we use a rather general notion, namely the expected distance between the real location and the reported location. This is a function of the prior distribution on the locations $\pi$, and of the conditional probabilities that determine the mechanism:
\begin{eqnarray}\label{eqn:utility}
\mathcal{U}(\Prob,\pi) = \sum_{x,y} \pi(x) \cdot\Prob(y\mid x) \cdot d(x,y)
\end{eqnarray}

As explained above, the optimal values  $\Prob(y\mid x)$ can be determined by solving a linear program with constraints \eqref{eqn:geo-ind} and objective function \eqref{eqn:utility}. Unfortunately, the number of the constraints  \eqref{eqn:geo-ind} is $\mathcal{O}(n^3)$, where $n$ is the number of locations. Hence, due to the complexity of linear programming, the method is unfeasible even when  $n$  is relatively small. 
To get an idea of the dimensions, consider the Quartier Latin in Paris, which has an area of about $1.5 \times 1.5$~km$^2$. If we set the size of the  locations to be $100 \times 100$~m$^2$, we need a grid of $15 \times 15 = 225$ cells to cover the area, which means $225^3$ constraints! 
Reducing the granularity of the grid (i.e., considering larger cells) is not a solution, because it degrades the meaning of the utility in \eqref{eqn:utility}, as discussed in \cite{Chatzikokolakis:17:POPETS}.
% Indeed, the distance $d(x,y)$ is to be interpreted as distance between the centers of the cells $x$ and $y$. 

%Due to space constraints, we omit some technical details. They can be found in the full version available at 

\section{Reducing the set of constraints}
We now propose a method to reduce the number of constraints of the linear program to  $\mathcal{O}(n^2)$, thus making the application of the method feasible for typical cases  like the above one. This will be at the price of some utility loss, i.e., our method will only approximate the optimal solution. We will see, however, that the loss is quite acceptable, while the gain in performance is significant.
 
Let $\mathcal{X}$ be the set of locations. Let $x_0, x_k\in \mathcal{X}$, and consider a path $x_1,\ldots, x_{k-1}\in \mathcal{X}$ from $x_0$ and $x_k$. Let $\delta$ be the smallest number such that
\begin{eqnarray}\label{eqn:maj}
\sum_{0}^{k-1}  d(x_i,x_{i+1})\leq \delta  \cdot d(x_0,x_k)
\end{eqnarray} 
Note that in general  $ \delta \geq 1$ because of the  triangular inequality.
It is easy to see that the constraint 
\begin{eqnarray}\label{eqn:consequence}
\Prob(y\mid x_0) \leq \exp(\varepsilon \cdot d(x_0,x_{k}))\cdot \Prob(y\mid x_{k})
\end{eqnarray}
is a consequence of all constraints of the form
\begin{eqnarray}\label{eqn:shrinked}
\Prob(y\mid x_i) \leq \exp(\nicefrac{\varepsilon \cdot d(x_i,x_{i+1})}{\delta}) \cdot \Prob(y\mid x_{i+1})  
%\Prob(y\mid x_i) \leq \exp\Big(\frac{\varepsilon \cdot d(x_i,x_{i+1})}{\delta}\Big) \cdot \Prob(y\mid x_{i+1})  
\end{eqnarray}
for $  i=0,\ldots, k-1$. Therefore, it is sufficient to consider a set of constraints  ${\mathcal{C}}$ of the form \eqref{eqn:shrinked}, containing enough elements so to deduce all original constraints of the form \eqref{eqn:geo-ind}. Namely, it is sufficient ensure that, for every   $x_0,x_h\in\mathcal{X}$, there are 
$x_1,\ldots, x_{h-1}\in \mathcal{X}$ such that all constraints  \eqref{eqn:shrinked} 
%\begin{eqnarray}\label{eqn:shrinkedC}
%\Prob(y\mid x'_i) \leq \exp( \nicefrac{\varepsilon \cdot d(x'_i,x'_{i+1})}{\delta}) \cdot \Prob(y\mid x'_{i+1}) 
%%\Prob(y\mid x'_i) \leq \exp\Big( \frac{\varepsilon \cdot d(x'_i,x'_{i+1})}{\delta}\Big) \cdot \Prob(y\mid x'_{i+1}) 
%\end{eqnarray}
are   in $\mathcal{C}$, for $ i=0,\ldots, h-1$. Then, to achieve  the original level $\varepsilon$ of indistinguishability (i.e., of location privacy), it is sufficient 
to solve a new linear program with the same objective function and set of constraints  $\mathcal{C}$. Note that in general the solution of the new program will give an utility inferior to the original one, because the constraints in $\mathcal{C}$ are stricter (i.e., enforce more privacy, due to the division by $\delta$) than the original constraints. 

We construct $\mathcal{C}$ as follows:  For every  $x \in \mathcal{X} $, we consider all  $x'\in \mathcal{X}$ such that $d(x,x')\leq R$, where $R$ is some fixed distance. Then, for every $y \in\mathcal{X}$, we add a constraint of the form  \eqref{eqn:shrinked}, where $x_i=x$ and $x_{i+1}=x'$.  
To make sure that we have enough elements in $\mathcal{C}$, we assume the following \emph{density hypothesis}: Let  $ch(\mathcal{X})$ be the \emph{convex hull} of $\mathcal{X}$ (represented as points in the map, for instance, the centers of the cells). Then:
\begin{eqnarray}\label{eqn:hyp}
\forall y \in ch(\mathcal{X}) \;\; \exists x \in\mathcal{X} : d(y,x) \leq \rho
\end{eqnarray}
where $\rho$ is some fixed distance.
Note that  to connect every pair of points in $\mathcal{X}$ it is necessary and sufficient to have $R\geq 2\rho$. 

The execution time depends on the cardinality of $\mathcal{C}$, which, for fixed $R$, is $\mathcal{O}(n^2)$ ($n$ being the number of locations in $\mathcal X$). The cardinality of $\mathcal{C}$ is also proportional to $R^2$, so from the point of view of efficiency it is convenient to keep $R$ as small as possible. 
On the other hand, the utility is monotonic on $R$, hence  the choice of $R$ must take into account the trade-off between efficiency and utility. 

\begin{figure}
%\centering
%\includegraphics[width=0.45\textwidth]{bound} 
\includegraphics[width=0.3\textwidth]{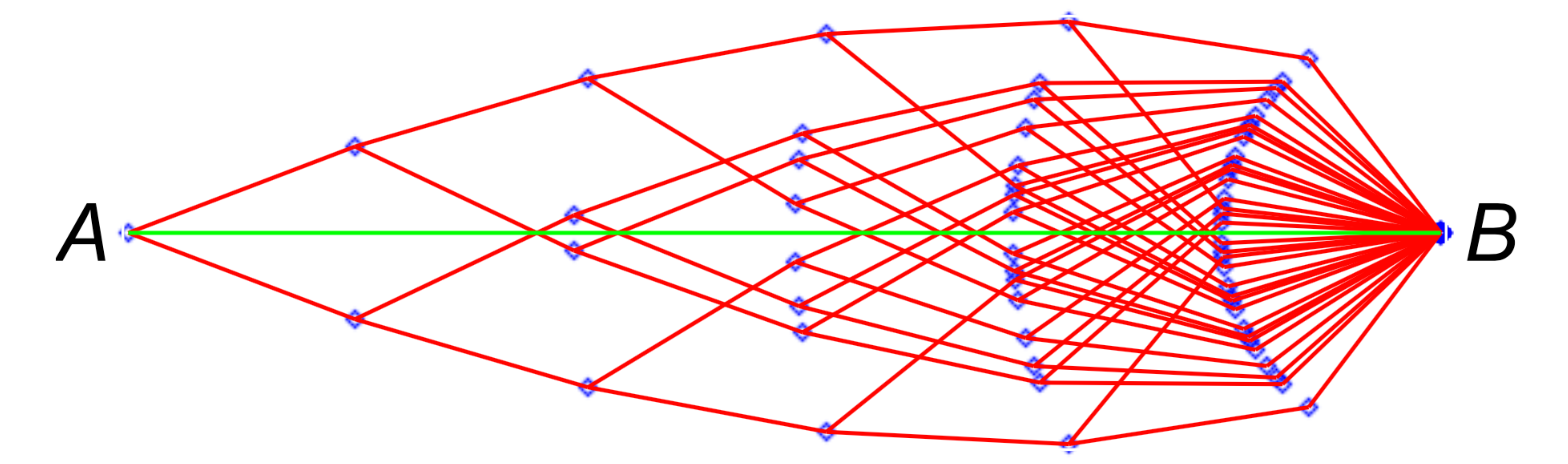} 
\vspace{-3mm}
\caption{
%\vspace{-3mm}
Possible paths between two locations $A$ and $B$. }
\label{fig:construction}
\end{figure}

\begin{figure}
%\centering
%\includegraphics[width=0.45\textwidth]{bound} 
\includegraphics[width=0.32\textwidth]{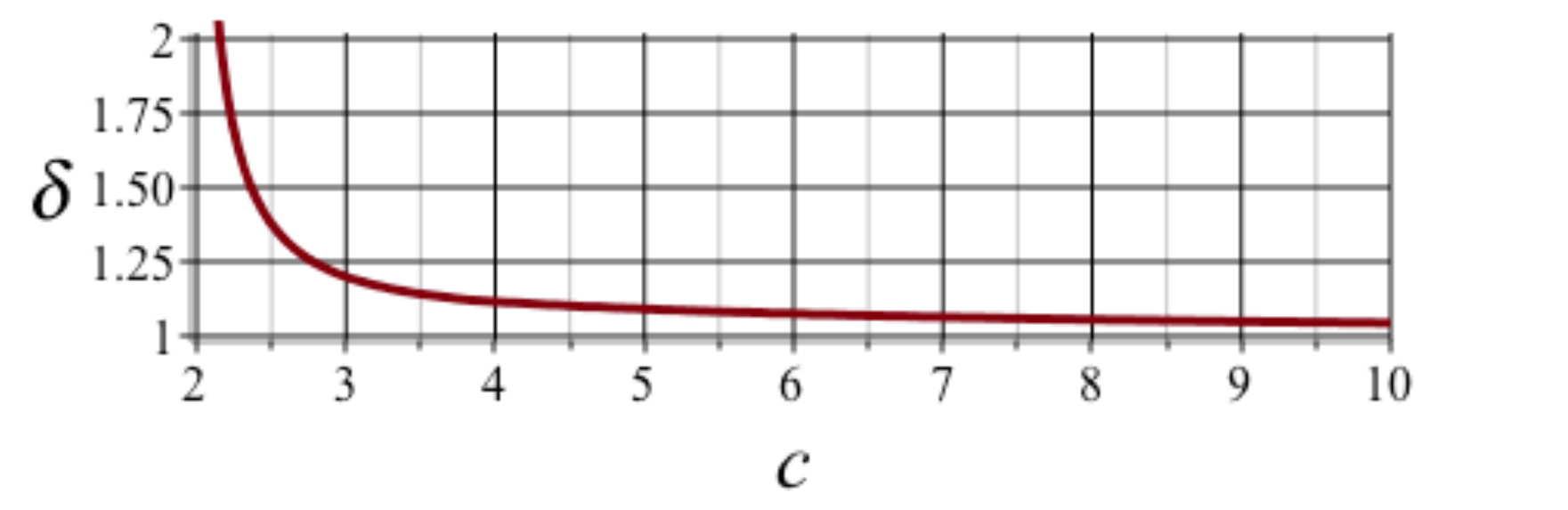} 
\vspace{-4mm}
\caption{
%\vspace{-3mm}
Worst-case value of $\delta$ as a function of $c$. }
\label{fig:bound}
\end{figure}

The utility loss depends  on  the expansion factor $\delta$ introduced in \eqref{eqn:utility}, which in turn depends on $R$ and $\rho$. 
%We are therefore interested in establishing a bound on $\delta$. 
%From geometrical considerations, we derive:  
%%\[
%%\delta \leq \frac{R (R - \rho)}{  \sqrt{R^4 - 2\rho R^3 - 3\rho^2 R^2 + 4\rho^3 R   + \rho^4} }
%%\]
%%or equivalently  
%%\[
%$\delta \leq \mathit{bound}(c)$, where 
%\[
%\mathit{bound}(c) = \frac{c(c-1)} {  \sqrt{c^4 - 2c^3 - 3c^2 + 4c   + 1}  }  \; \; \mbox{and}\; \; c=\frac{R}{\rho}
%\]
%The bound is rather loose for small values of $c$, in particular in the proximity of $2.5$, but then as $c$ grows it approaches rapidly  the ideal value $1$. For instance,  we have  $\mathit{bound}(3)=1.66$ and $\mathit{bound}(5)=1.11$. 
The analysis of the worst-case for $\delta$ can be done by the geometrical construction illustrated in Figure~\ref{fig:construction}. Given two locations $A$ and $B$, the lines in red represent the possible paths between 
$A$ and $B$  guaranteed by the density hypothesis~\eqref{eqn:hyp}.    
%\vspace{-3mm}
Figure~\ref{fig:bound} shows the graph of the worst-case $\delta$ as a function of the ratio $c=\nicefrac{R}{\rho}$. 
(Due to the condition $R\geq 2\rho$, $\delta$ is not defined for $c<2$.) We note that $\delta$ becomes very high when $c$ is close to $2$, but it approximates rapidly the ideal value $1$ as $c$ grows.  
%\vspace{-3mm}

\section{Evaluation}
In this section we evaluate our method and compare with the optimal approach. We consider a set of locations disposed along the intersection points of a grid, and we set the distance $u$ between two adjacent locations as the unit distance, i.e.,  all distances will be expressed in terms of $u$. The results illustrated in this section  are valid for any  value of $u$. For the example of the Quartier Latin, for instance, we could consider $u= 100$ m. 

We note that in such grid of locations, $\rho = \nicefrac{1}{\sqrt{2}}$. Indeed, the points at maximum distance from any location are the centers of the cells, which are at distance $\nicefrac{1}{\sqrt{2}}$ from the corners of their cell. Concerning the prior, we  consider a uniform distribution. We also fix $\varepsilon = \nicefrac{\ln 2}{2}$, which means a level of indistinguishability $2$ in a radius of $2u$.  For instance, for $u=100$ m,  
a user would have protection $2$ in a radius of $200$ m, i.e., 
from the point of view of an adversary, the user's real location could be no more than  twice more likely than any location within $200$ m from it.    

\begin{figure}[h!]
\centering
\includegraphics[width=0.485\textwidth]{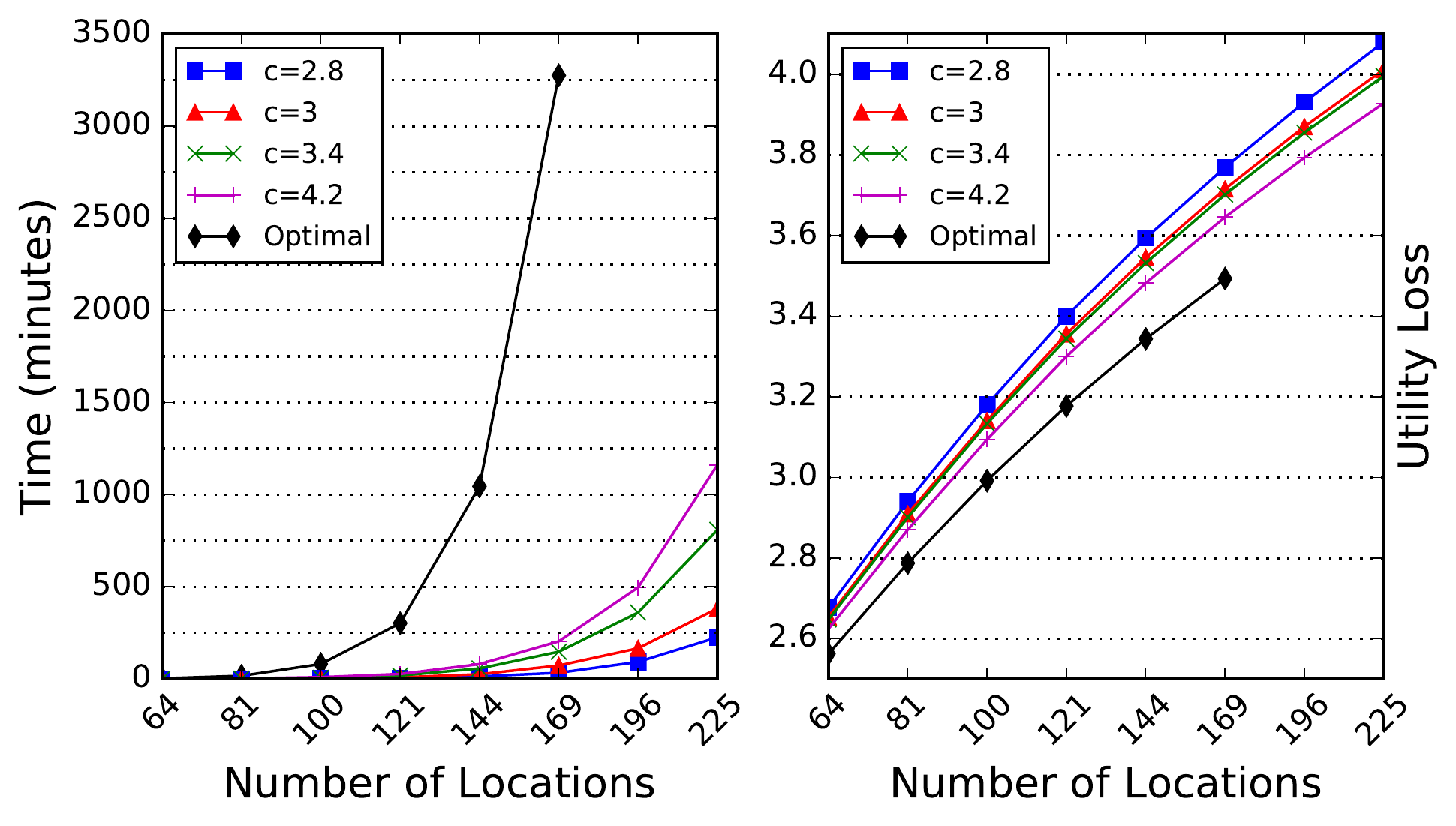} 
\vspace{-6mm}
\caption{
Execution time and utility loss.
% for both our method and optimal one, for different number of locations and  values of $R$. 
 }
\label{fig:results}
\end{figure}

We  experimented with grids from $8\times 8=64$ up to $15\times 15=225$ locations, and values of $c$  from $2.8$ ($R= 1.98$) to $4.2$ ($R= 2.97$), using an Intel machine (no TSX) 2VCPUs 2.3 GHz, 4GB RAM.  
 The resulting computation times and   utilities are shown in Figure~\ref{fig:results}. We did not evaluate the performance of the optimal method for more than  $169$ locations because it was taking too much time:
% . For instance,  
 with $169$ locations it took $3,275$ minutes  (more than $2$ days), 
 and 
 with $196$ locations it was still running after several days. 
 
We can see that with $225$ locations like in the example of the Quartier Latin, the optimal method would be completely unfeasible. With our method  and  $c=2.8$ it takes $226$ minutes (note that this computation is just to \emph{build} the mechanism, so it is  done only once; afterwards, the use of the mechanism is immediate), and the utility loss is not much higher than that of the optimal method. For instance, on $169$ locations, they are $3.77$ and $3.49$, respectively. 

%
%\vspace{-3mm}

\section{Future work}
As future work, we plan to improve the bound on $\delta$,  use a prior based on real location data, and  compare our method also with the one based on Laplacian noise and remapping  proposed in \cite{Chatzikokolakis:17:POPETS}. 

\bibliographystyle{ACM-Reference-Format}
\bibliography{short} 

\end{document}